\documentclass[twocolumn]{aa}
\usepackage{graphics,natbib,amssymb}
\usepackage{rotating}
\usepackage{graphicx}
\usepackage{txfonts}
\newcommand{\bm}[1]{\mbox{\boldmath $#1$}}
\newcommand{\bM}[1]{\mbox{\textbf{\textsf{#1}}}}
\begin{document}
   \title{A new inverse method for stellar population synthesis and error analysis}

   \subtitle{}

   \author{J. Moultaka
          \inst{1,} \inst{2}}


   \institute{Physikalisches Institut, Z\"ulpicher Str. 77, 50937 K\"oln\\
           \email{moultaka@ph1.uni-koeln.de}
         \and
            LUTH, Observatoire de Meudon,
           5, place Jules Janssen,
           92190 Meudon cedex, France \\ 
             }

   \date{Received ; accepted }

   \abstract{The stellar population synthesis in unresolved composite objects is a very tricky problem. Indeed, it is a degenerate problem since many parameters affect the observables. The stellar population synthesis issue thus deserves a deep and rigourous analysis. In this paper we present a method of inversion which uses as observables the intensities at each pixel of a galactic spectrum and provides the stellar contribution to luminosity of all stars considered in a database. The main contribution of this paper to the synthesis problem is that it provides an analytical computation of the uncertainties accompanying a solution. This constitutes an important improvement relative to previous methods which do not provide such infomation except in the method described by Pelat (1997) and Moultaka \& Pelat (2000). The latter uses the equivalent widths and intensities of stellar spectra in order to reproduce the equivalent widths of a galactic spectrum. The novelty of this work relative to the previous one is that the dust emission present in the IR spectra can be modeled as well as the velocity dispersion of stars that broadens the lines of a galactic spectrum. Tests are also performed in order to estimate the reliability of the method and the influence on the results of an additive continuum present in a studied spectrum, for example in the case of AGNs.   
   }

   \maketitle
%

\section{Introduction}\label{sec:intro}
Thanks to the analysis made by Scheiner (1899) and later by Ohman (1934) who noticed the similarity between the spectra of the sun and the nucleus of M32, the idea of considering a galactic spectrum as the product of a combination of stellar spectra was born. Since that time, different works were carried out with the aim of providing the stellar populations of composite objects by means of stellar spectra.\\ 
The main problem of the stellar population synthesis principle is that it is a 
degenerate problem. A large number of free parameters contribute to the making of the observed composite object spectrum (a galaxy or a cluster). 
Examples of such parameters are the stellar luminosity classes, stellar spectral types, metallicities, the physical characteristics and amount of dust, the geometry of the system, the presence of an AGN etc... 
All these parameters differently affect the observed spectrum of a composite object according to the considered wavelength domain of the spectrum (e.g. the production of absorption lines, emission lines, additive featureless continuum, reddening etc...).\\
In order to deduce the stellar population of a galaxy or a cluster, two approaches have been adopted, the ``direct'' approach and the ``inverse'' one. In the direct approach (e.g. Tinsley 1972, Charlot \& Bruzual 1991, Bruzual \& Charlot 1993, Leitherer et al. 1999, Fioc \& Rocca-Volmerange 1997, Vazdekis, 1999, Bruzual \& Charlot 2003), an {\em a priori} model of the studied object is constructed by choosing the shape of the initial mass function, the star formation rate, the geometry, the rate of supernovae and/or any other parameters. As the system evolves with time, synthetic observables are constructed at each time step with the use of stellar evolutionary models (the ``Padova'' and ``Geneva'' groups, e.g. Girardi et al. 2004, Charbonnel et al. 1999 and references  herein) and directly compared to the real observables at all time steps. The retained solution is the one that best reproduces the observables. \\
This approach makes it very hard to estimate the uncertainties on the stellar population and all the free parameters. \\
In the inverse approach (e.g. Faber 1972, Joly 1974, O'Connell 1976, Bica 1988, Schmidt et al. 1989, Silva 1991, Pelat 1997,1998, Moultaka \& Pelat 2000) one does not assume any predefined model for the stellar composition of the studied object. The observables and a method of inversion are only used in order to derive the characteristics of the stellar population.\\
This is done by minimising a function called the objective function; it represents the goodness of the fit between the synthetic model and the observables. The estimation of the error bars on the stellar population is more confortable in this approach.
However, it is usually made using Monte-Carlo simulations which require a large number of simulating spectra. The synthesis procedure becomes thus quickly very costly in computing time and provides biased results. \\

An inverse method described in Pelat (1997), Moultaka \& Pelat (2000) and Moultaka et al. (2004) deals with the problem of minimisation in the sense that it provides a unique global minimum (Pelat 1997). It provides as well an analytical error analysis (Moultaka \& Pelat 2000) and a study of the influence of constraints added {\em a priori} while searching for a solution (Moultaka et al. 2004). The method uses the equivalent widths (EWs) of the absorption lines of a galactic (or any composite object) spectrum as observables to be fitted by a combination of EWs of the same absorption lines and continuum intensities of stellar spectra.\\
This method is limited as it requires the measurement of equivalent widths and consequently, an assumption of the location of the continuum in the spectra. Moreover, at high spectral resolution, the equivalent widths measurements become very problematic as one needs to distinguish all available lines and determine their intervals. On the other hand, the dust emission occuring particularly in the IR part of the spectrum is not modeled, nor is the velocity dispersion of stars which broadens the absorption lines and becomes very effective at high spectral resolution. \\

In this paper, we present a new method of inversion which uses all spectral pixel intensities of a spectrum as observables instead of equivalent widths of absorption lines and we provide an analytical error analysis. The reddening in the optical domain due to the presence of dust, the dust emission in the IR and the velocity dispersion of stars are also modeled.\\
The method and the error analysis are described in the next section and tested in Sect. 4. In Sect. 3, we discuss the properties of the database used in the synthesis problem. An application to the globular cluster G 170 is described in Sec. 5. In appendices B and C are listed respectively the facilities of the code and the output files.  


\section{The inverse method}\label{sec:method}

\subsection{Search for the solution :}\label{sec:searchsolution}
First, we consider a simple case where a galaxy (or a composite object) is only composed of stars. This assumption leads us to consider the spectrum of a galaxy as the result of the sum of stellar spectra. Each pixel intensity of the galactic spectrum is then obtained as follows:
\begin{equation}
F_{\lambda gal}=\sum_{i=1}^{n_{\star}}F_{\lambda i}
\label{Flambdagal}
\end{equation}
where $F_{\lambda gal}$ is the galactic intensity at wavelength $\lambda$ and $F_{\lambda i}$ the intensity at the same wavelength of a star of {\it class} $i$ (i.e. spectral type and luminosity class) of the stellar database. $n_{\star}$ is the number of stars considered in the database.\\

Let us call $k_{\lambda i}$ the contribution to the galactic luminosity at a wavelength $\lambda$ of stars of class i (i.e. $k_{\lambda i}= \frac{F_{\lambda i}}{F_{\lambda gal}}$). If we make the assumption that the galaxy is only composed of stars considered in the database then :
\begin{equation}
\sum_{i=1}^{n_{\star}} k_{\lambda\,i}=1
\label{eq:sum1}
\end{equation} 
At a reference wavelength $\lambda_0$ one can write $k_{\lambda_0\,i}= \frac{F_{\lambda_0\,i}}{F_{\lambda_0\,gal}}$ which implies that:
\begin{equation}
\frac{F_{\lambda gal}}{F_{\lambda_0 gal}}=\sum_{i=1}^{n_\star} k_{\lambda_{0\,i}}\frac{F_{\lambda i}}{F_{\lambda_{0i}}}
\end{equation}
or equivalently $I_{gal \,j}=\sum_{i=1}^{n_\star} k_i I_{ji}$, where $k_i=k_{\lambda_{0\,i}}$ and $I_{gal \,j}$ and $I_{ji}$ are respectively the galactic and class $i$ stellar intensities at wavelength $\lambda$ (indexed with $j$) normalised to the intensity at $\lambda_0$ (i.e. $I_{gal \,j}=\frac{F_{\lambda gal}}{F_{\lambda_0}}$ and $I_{ji}=\frac{F_{\lambda i}}{F_{\lambda_0}}$).\\

A synthetic spectrum can then be constructed in a similar way (equivalent to Eq. (\ref{Flambdagal})) as follows:
\begin{equation}
F_{\lambda syn}=\sum_{i=1}^{n_{\star}}F_{\lambda i}
\label{equ:fsyn}
\end{equation}
where $F_{\lambda syn}$ is the synthetic intensity at wavelegnth $\lambda$. Using similar notations as previously, we obtain the synthesis equation:
\begin{equation}
I_{syn \,j}=\sum_{i=1}^{n_\star} k_i I_{ji}
\label{equ:Isyn}
\end{equation}  
In order to ensure that a solution is physically acceptable, the stellar contributions to luminosity $k_i$ have to be all positive. Thus, considering in addition the condition of Eq.~(\ref{eq:sum1}), the synthesis problem is finally stated as follows:
\begin{equation}
\left\{
\begin{array}{ll}
\bm {II} \bm k =\bm {I_{syn}} &= \bm {I_{gal}}\\
\bm {k} & > 0\\
\bm {b} \bm k&=1
\end{array}
\right.
\label{eq:systeme}
\end{equation}
In the previous system, $\bm {II}$ is the $n_l$x$n_{\star}$ matrix composed of the $n_l$ normalised intensities at each pixel of the $n_{\star}$ stellar spectra (i.e. composed of the $I_{ji}$), \bm {I_{gal}} and \bm {I_{syn}} are respectively the vectors composed of the $n_l$ galactic and synthetic normalised intensities at each pixel (i.e. respectiveley $I_{gal\, j}$ and $I_{syn\, j}$). Finally, $\bm k$ is the vector composed of the $n_{\star}$ stellar contributions to luminosity $k_i$.\\
Solving the synthesis problem becomes thus to find the set $\bm k$ of stellar contributions to luminosity at the reference wavelength that minimises the following objective function:
\begin{equation}
f(\bm k)=\|\bm {I_{gal}}-\bm {II} \bm k\|^2  
\end{equation}
This expression is equivalent to the ``synthetic distance'' defined by Pelat (1997) for the EW case and represents the ``mean residual intensities''. We keep the same terminology and express the synthetic distance in our case:
\begin{equation}
\begin{array}{ll}
D^2=f(\bm k)&=\sum_{j=1}^{n_l}(I_{gal\,j}-\sum_{i=1}^{n_{\star}} k_i I_{ji})^2 \\
   &=\sum_{j=1}^{n_l}(I_{gal\,j}-I_{syn\,j})^2
\end{array}\label{eq:distance}
\end{equation}
As shown in Eq.~(\ref{eq:distance}), the synthetic distance for the case of intensities happens to be similar to the ``principal synthetic distance'' defined by Pelat (1997) for the case of EWs. Therefore, the problem is simplified in this case and possesses a unique solution. It is the solution of the least square problem subject to the contraints shown in equation system (\ref{eq:systeme}).  

The minimisation procedure is done using the constrained least square method of type LS1 of the NAG library.

\subsection{Reddening, dust emission and velocity dispersion:}
Due to the presence of dust in the Milky Way and inside the studied galaxy, a galactic spectrum is reddened in the optical. This is due to the absorption by dust of the light which is higher in the blue than in the red part of the spectrum. \\
The intensities of a reddened spectrum are transformed as follows:
\begin{equation}
F_{reddened}(\lambda)=F(\lambda)10^{(-0.4*F_{Law}(\lambda)E(B-V))}
\end{equation}
where $F(\lambda)$ is the intensity at wavelength $\lambda$, $F_{reddened}(\lambda)$ the intensity at the same wavelength of the reddened spectrum and $F_{Law}(\lambda)$ the reddening law. The latter can be chosen and included in the code very easily (see Appendix B). The present law used for the synthesis in Sects. 3 and 4 is Howarth's law (Howarth 1983).\\
In the present synthesis method, the insertion of the reddening effect into an optical spectrum is made by constructing different databases of stellar spectra reddened by a certain amount of E(B-V) and performing the synthesis using the different databases. The solution providing the smallest synthetic distance is retained with the corresponding value of E(B-V). \\

A second effect due to the presence of dust is an additive emission produced by the heated dust. It can be described by a blackbody continuum of low temperature (T$\sim 500K-1000K$) and therefore, is assumed to be negligible in the optical domain.\\
Dust emission in the IR case is included as an additional star. The search for a solution is made by adding to the database a blackbody continuum with a given temperature (or any other chosen function, see Appendix B) and performing the inversion. The test is repeated for different temperatures of the blackbody and, as for the reddening, the retained solution is the one which minimises the synthetic distance. The corresponding dust temperature is adopted for the model.\\

The velocity dispersion of stars in a galaxy influences the spectrum of the latter by a broadening effect of the stellar absorption lines. This effect obviously influences the results especially at high spectral resolution. \\
The velocity dispersion is included in the code by testing the results obtained with different stellar databases constructed from the initial database by smoothing all the spectra with a given value of velocity dispersion. The convolution is made with a Gaussian function but any other function can be inserted easily in the code (see Appendix B).

\subsection{Error analysis :}\label{sec:errors}
In the following, we assume that the contribution to the errors from the spectra of the database is negligible compared to that of the galaxy. If this would not be the case, there would be no sense in performing the synthesis with such bad data.

Given an approximate solution $\bm I_{syn \,0}$, equation system~(\ref{eq:systeme}) is still valid when $\bm I_{gal}$ is replaced by $\bm I_{syn \,0}$.\\ 
After differentiating the obtained equations, one gets the following system:
\begin{equation}
\left\{
\begin{array}{ll}
\bm {II} d\bm k &= d\bm {I_{syn}} \\
\bm {b} d\bm k&=0
\end{array}
\right.
\label{eq:systemedif}
\end{equation}
As the synthetic distance is a Euclidean distance, one can write $d\bm {I_{syn}}=\bM H d\bm {I_{gal}}$ where $\bM H$ is the orthogonal projector on the hyperplane tangent to the ``synthetic surface'' (i.e. the set of intensities having an exact solution, see Appendix A and Pelat 1997) at $\bm I_{syn\,0}$. It follows from equation system (\ref{eq:systemedif}):  
\begin{equation}
\left\{
\begin{array}{ll}
\bm {II} d\bm k &= \bM H d\bm {I_{gal}}\\
\bm {b} d\bm k&=0
\end{array}
\right.
\end{equation}
This translates into:
\begin{equation}
d\bm k=(\bm {II}^{+\,t}\bm {II}^{+})^{-1}\bm {II}^{+\,t}\left(\begin{array}{c}
\bM H d\bm I_{gal}\\
\bm 0
\end{array}\right)
\label{eq:dk}
\end{equation}
where  $\bm {II}^{+}=\left(\begin{array}{c}\bm {II}\\
                                                      \bm b 
\end{array}\right)$.\\
The variance-covariance matrix of the stellar contributions ($\bM V_k$) can be deduced from Eq.~(\ref{eq:dk}) and written as a function of the variance-covariance matrix of the galactic normalised intensities $\bM V_I$ as follows :
\begin{equation}
\begin{array}{ll}
\bM V_k &=<d\bm k,d\bm k^t> \\
        &= (\bm {II}^{+\,t}\bm {II}^{+})^{-1}\bm {II}^{+\,t}\bM H <d\bm I^{+}_{gal},d\bm I^{+\,t}_{gal}> \bM {H}^t{II}^{+}(\bm {II}^{+\,t}\bm {II}^{+})^{-1}\\
        &= (\bm {II}^{+\,t}\bm {II}^{+})^{-1}\bm {II}^{+\,t}
\left(\begin{array}{cc}
\bM H \bM V_I \bM {H}^t & \bm 0\\
\bm 0& 0
\end{array}\right)
{II}^{+}(\bm {II}^{+\,t}\bm {II}^{+})^{-1 t}
\end{array}
\end{equation}
where $d\bm I^{+}_{gal}=\left(\begin{array}{c}d\bm I_{gal}\\
                                                      0 
\end{array}\right) $.\\

In order to compute the orthogonal projector $\bM H$, the following two methods can be used: 
\begin{itemize}
\item From the first equation of equation system (\ref{eq:systemedif}) ($\bm {II} d\bm k = d\bm {I_{syn}}$) it follows that the vector of the deviation around the synthetic normalised intensities $d\bm {I_{syn}}$ belongs to the linear envelope of the columns of matrix $\bm {II}$. This implies that the linearly independent column vectors of this matrix form a basis of the tangent hyperplane when the origin of the vector space of the normalised intensities is shifted to $\bm {I_{syn \,0}}$ (because, in this case, d\bm {I_{syn}} would be on the tangent plane). A QR decomposition of matrix  $\bm {II}$ (e.g. Golub \& Van Loan 1996) allows one to build the projector $\bM H$.

\item The second way of computing the orthogonal projector $\bM H$ is to search for the equations describing the synthetic surface ($Q_r(I_{syn\,1},I_{syn\,2},...,I_{syn\,r})=0$) as well as the gradients at $\bm I_{syn \,0}$ of functions $Q_r$ relative to $I_{syn\,j}$. These gradients are then arranged in column vectors of a matrix $\bM G$. As in the previous method, the orthogonal projector is deduced from a QR decomposition of matrix $\bM G$.\\
The search for functions $Q_r$ and their gradients is described in Appendix A. 
\end{itemize}

Using the orthogonal projector $\bM H$, one can also compute the error on the synthetic distance in the same way as described in Moultaka et al. (2004) for the EW case. A similar result is obtained in the case of intensities :
\begin{equation}
\sigma_{D^2}=2\sqrt{(\bm I_{gal \,0}-\bm I_{syn \,0})^T(\bM H-\bM Id)\bM V_I(\bM H^T-\bM Id)(\bm I_{gal \,0}-\bm I_{syn \,0})}
\end{equation}

The error bars on the reddening and velocity dispersion are estimated using the different values of these parameters tested in each run and providing synthetic distances which lie inside the error bars of the minimal distance.

\subsection{Constraining the solution}\label{sect:constraints}

The solution provided by the minimisation procedure is, in mathematical terms, the best solution. However, this solution may not satisfy (astro)physical constraints which are not taken into account in Sect. \ref{sec:method} (except the positivity constraint). For this reason, it is crucial to constrain the solution in such a way that it is (astro)physically acceptable. \\
This is done by minimising, as previously, the synthetic distance defined in Eq.~(\ref{eq:distance}) while the contributions to luminosity at the reference wavelength $\bm k$ are subject to linear constraints written as follows:
\begin{equation}
\bm l\le
\left\{
\begin{array}{c}
\bm k\\
\bM C\bm k
\end{array}\right\}\le \bm u
\label{eq:contr}
\end{equation}
In the previous equation, $\bm l$ and $\bm u$ are respectively lower and upper constant vector limits, $\bM C$ is the constraint matrix composed of the linear coefficients of the stellar contributions to luminosity $k_i$. In order to take into account the second constraint shown in equation system~(\ref{eq:systeme}) ($\bm {bk}=1$), the first line of matrix $\bM C$ is composed of a series of $1$ and the second components of vectors $\bm l$ and $\bm u$ are equal to $1$. The positivity constraint is included in Eq.~(\ref{eq:contr}) if the first components of vectors $\bm l$ and $\bm u$ are respectively equal to zero and infinity.\\  
Note that in the present case (the case of intensities), one does not need to iterate on the synthetic surface as is done in Moultaka et al. (2004) where the observables are the EWs of the absorption lines, because the synthetic distance here is exaclty the ``principal synthetic distance''. \\
As is described in Moultaka et al. (2004), the constraints are chosen in such a way that the Initial Mass Function (IMF) of each episode of star formation in the studied galaxy satisfies the standard IMFs known in the litterature.\\
Two types of constraints can be used in the present version of the method: the ``Decreasing IMF mode'' where the only hypothesis on the IMF is that it is a decreasing function. The second type of constraint is the ``Standard mode'' where the constraints on the IMF are generalised in such a way that the standard IMFs (Salpeter 1955, Scalo 1986, Kroupa et al. 1993) are all satisfied. The constraints are then imposed on the numbers (or equivalently on the contributions to luminosity) of dwarf stars on the one hand and on dwarf and more evolved stars (supergiants included) on the other hand. A detailed description of the constraint model is given in Moultaka et al. (2004).\\

The facilities of the code and the description of the output files ar listed in Appendices B and C.

\section{Properties of the database}\label{sec:database}
The choice of the stellar (or cluster) database is of great importance. Indeed, the solution of a synthesis depends closely on this choice and consequently, the interpretation of the stellar population of a composite object may be biased. Therefore, a database should be the most informative available. This means that the wavelength domain should be the most extended, the spectral resolution the highest, and the number of stars (or clusters) the largest possible, including all stellar classes (i.e. spectral types, luminosity classes and metallicities) in the case of stars or a large sample of ages and metallicities in the case of clusters. However, the spectral range and spectral resolution are finite and must be comparable to those of the composite object spectrum, which restricts the information extracted from the spectra. Consequently, the different components of the database are less distinguished and their number can hardly be as large as the variety of known stellar (or cluster) classes. 
For this reason, at a given spectral range and resolution, a database should be built in such a way that its different components are not linearly correlated with each other or equivalently in our case, that no spectrum of the database can be synthesised by the remaining spectra. Indeed, if two spectra were not distinguished in the sense mentioned above and contribute to the synthesis solution of a composite object, it would be impossible to assert whether both stellar classes are present in the stellar population of the composite object or only one of them. Consequently, the number of stars (or clusters) in the database should decrease with the spectral resolution or the spectral range.\\
In addition, as it has been shown in Moultaka \& Pelat (2000) (see their Fig. 2), the choice of the database has to be adapted to the quality of the composite object observation i.e. if two uncorrelated stellar (or cluster) spectra were separated from each other within the noise level of the composite object spectrum (the S/N ratio of the stellar (or cluster) spectra is assumed to be always much higher than that of the composite object spectrum, see Sect. \ref{sec:errors}), they should not be both kept in the database because the quality of the composite object spectrum would not allow us to distinguish between them.\\
Moreover, it is obvious that the relative flux calibration of the stellar (or cluster) spectra is absolutely necessary and crucial in this method as the solutions depend closely on the different relative fluxes of each spectrum (see Eq. \ref{equ:Isyn}). However, the absolute flux calibration is not needed as the spectra are all normalised to one at the reference wavelength (see Sect. \ref{sec:searchsolution}) but the magnitudes of the stellar classes would be necessary if one wishes to derive the number of stars from the stellar contributions to luminosity ($\bm k_i$) and constrain the solutions as described in Sect.~\ref{sect:constraints}.

\section{Test of the method}\label{sec:test}
A normal test of the method is to simulate a galaxy by choosing different contributions $k_i$ of the stars as an input ($\bm k_{Input}$) and construct the spectrum of a galaxy with the given contributions and the stellar spectra. The inversion of the simulated galaxy with the method described in Sect.~\ref{sec:method} will provide an estimate of the efficiency of the method for different values of the signal to noise ratio of the spectra.  \\
We have chosen 11 stars out of the 31 stars of Boisson et al. (2000) in order to perform the simulation. The 11 stars have been chosen in such a way that the different stellar classes (hot and cool stars, dwarfs, giants and supergiants, solar metallicity and SMR stars) with different continuum slopes are represented best. The spectral resolution of this database is of $11\AA$. The choice of this database is motivated by the fact that a higher number of stars would considerably increase the computing time for each simulation. Moreover, as our aim is to have a good representation of the different stellar classes, the spectral resolution of this database is sufficient to perform the simulations.\\
The simulated galaxy has been constructed after convolving the stellar spectra with a Gaussian distribution of parameters $\mu =0$ and $\sigma =140$ km/s. The obtained spectrum is then reddened by $E(B-V)=0.2$ mag using the reddening law of Howarth (1983).

\subsection{Simulating the errors}
Considering the Signal to Noise ratio ($S/N$) at each pixel as the ratio of the mean value of the normalised intensity $F_{mean}$ over its standard deviation $\sigma_F$, the standard deviation of the normalised intensity at each pixel is obtained for each S/N ratio through the relation:
\begin{equation}
\sigma_F=\frac{F_{mean}}{S/N}
\label{sigresol}
\end{equation}

We apply to the normalised intensities at each pixel of the simulated galaxy a random error following a normal law of parameters $\mu$ equal to the simulated value $F_{mean}$ and $\sigma$ to the standard deviation calculated in Eq.~(\ref{sigresol}) for a given spectral resolution.  \\
Monte-Carlo simulations are then performed in order to determine the mean values of the stellar contributions, the reddening and the velocity dispersion as well as their standard deviations. 

\begin{table*}[htbp]
\begin{center}
\begin{tabular}{|r|c|c|c|c|c|}
\hline
star & $\bm k_{Input} (\%)$ & $k_{Output} (\%)$ & $k_{Output} (\%)$ & $k_{Output} (\%)$ & $k_{Output} (\%)$ \\
     &                 & $S/N=\infty$   & $S/N=100$    & $S/N=50$     & $S/N=25$\\
\hline
O7-B0V & 1.51 &  1.51 & $1.46\pm 1.52$ &$1.55\pm 2.11$  &$1.67\pm3.09$\\
A1-3V & 3.03  &  3.03  & $2.57\pm 2.33$ &$1.85\pm 2.74$  &$1.87 \pm3.58$\\
F2V & 4.54 &  4.54 & $4.97\pm 5.12$ & $5.90\pm 7.04$  &$6.19\pm8.56$\\
rG5IV & 6.06 &  6.06  & $6.25\pm 4.00$ &$6.58\pm 6.06$  &$6.04 \pm7.85$\\
rK3V & 7.58 &  7.58   & $7.31\pm 1.97$ &$6.72\pm 3.56$  & $6.36\pm5.13$\\
rM1V & 9.09 &  9.09  & $9.07\pm 0.51$  &$9.08\pm 1.02$  & $9.23\pm2.09$\\
G9III & 10.61 &  10.61& $11.08\pm 5.85$ &$12.07\pm 9.99$ &$14.26 \pm15.89$\\
rK3III & 12.12 &  12.12 & $11.90\pm 3.52$ &$12.09\pm6.51$ & $13.59 \pm10.07$\\
M5III & 13.64 &  13.63 & $13.64\pm 0.19$ &$13.63\pm0.36$ &$13.68 \pm0.71$\\
rG2I & 15.15 &  15.15 & $15.15\pm 3.87$ &$14.12\pm7.38$ & $11.26\pm11.35$\\
M2I  & 16.67 &  16.67 & $16.59\pm 0.79$ &$16.41\pm1.55$ &$15.86\pm3.54$\\
\hline
$D^2$ & -        &  $2\,10^{-15}$  & $0.175\pm 0.009$ & $0.702\pm0.040$ & $2.812\pm0.172$    \\
\hline
E(B-V) & 0.2 & 0.2 &$0.20\pm  0.005 $ & $ 0.20\pm 0.01$ & $0.20\pm 0.02$ \\
\hline
$\sigma_{Dispersion}(km/s)$ &140  & 140 &$140.2\pm6.0$ & $140.6\pm 11.9$ & $137.0\pm 26.0$\\
\hline
\end{tabular}
\end{center}
\caption{Synthesis of a simulated galaxy constructed with the contributions to luminosity listed in the second column. The stellar population is dominated by intermediate- and late-type stars. The different solutions correspond to different values of the S/N ratio ($\infty$, 100, 50 and 25). $D^2$ is the synthetic distance, E(B-V) the reddening and $\sigma$ the dispersion.}
\label{tab:sim1}
\end{table*}

\begin{table*}[htbp]
\begin{center}
\begin{tabular}{|r|c|c|c|c|c|}
\hline
star & $\bm k_{Input} (\%)$ & $k_{Output} (\%)$ & $k_{Output} (\%)$ & $k_{Output} (\%)$ & $k_{Output} (\%)$ \\
     &                 & $S/N=\infty$   & $S/N=100$    & $S/N=50$     & $S/N=25$\\
\hline
O7-B0V & 16.67 & 16.67  & $16.38\pm2.09 $ &$15.76\pm4.19 $  &$13.34\pm7.57$\\
A1-3V & 15.15  & 15.15   & $15.69\pm4.05 $ &$15.55\pm6.34 $  &$15.74 \pm9.48$\\
F2V & 13.64 &  13.64  & $12.66\pm7.47 $ & $12.75\pm10.82 $  &$12.14\pm13.14$\\
rG5IV & 12.12 & 12.12   & $12.16\pm4.02 $ &$11.68\pm6.98$  &$ 9.94\pm9.41$\\
rK3V & 10.61 & 10.61    & $10.39\pm1.54 $ &$9.79\pm3.08 $  & $8.52\pm5.22$\\
rM1V & 9.09 &  9.09  & $9.07\pm0.35 $  &$9.12\pm0.73 $  & $9.31\pm1.53$\\
G9III & 7.58 & 7.58 & $8.58\pm6.05 $ &$9.95\pm9.43 $ &$ 13.39\pm14.55$\\
rK3III & 6.06 & 6.06  & $5.75\pm3.42 $ &$5.97\pm5.30$ & $ 5.77\pm6.61$\\
M5III & 4.54 & 4.54  & $4.56\pm0.11 $ &$4.59\pm0.21$ &$ 4.71\pm0.51$\\
rG2I & 3.03 & 3.03  & $3.29\pm2.56 $ &$3.48\pm4.16$ & $5.47\pm7.73$\\
M2I  & 1.51 & 1.51  & $1.45\pm0.52 $ &$1.35\pm0.95$ &$1.66\pm1.92$\\
\hline
$D^2$ & -        &  $5\,10^{-16}$  & $0.008\pm 0.004$ & $0.340\pm0.019$ & $1.359\pm0.076$    \\
\hline
E(B-V) & 0.2 & 0.2 &$0.20\pm  0.01 $ & $ 0.20\pm 0.02$ & $0.19\pm0.05$ \\
\hline
$\sigma_{Dispersion}(km/s)$ &140  & 140 &$141.4\pm9.1$ & $139.0\pm 21.9$ & $139.2\pm 42.9$\\
\hline
\end{tabular}
\end{center}
\caption{Same as in Table~\ref{tab:sim1}, for a population dominated by young stars.}
\label{tab:sim2}
\end{table*}

\begin{table*}[htbp]
\begin{center}
\begin{tabular}{|r|c|c|c|c|c|}
\hline
star & $\bm k_{Input} (\%)$ & $k_{Output} (\%)$ & $k_{Output} (\%)$ & $k_{Output} (\%)$ & $k_{Output} (\%)$ \\
     &                 & $S/N=\infty$   & $S/N=100$    & $S/N=50$     & $S/N=25$\\
\hline
O7-B0V & 6.06 & $6.06 $ & $5.76\pm 2.08$  & $5.42\pm 3.26 $ & $ 5.31\pm 4.81$ \\
A1-3V & 7.58  & $7.58 $ & $7.86\pm 3.80$ & $7.63\pm 5.30$ & $ 8.73\pm 7.42$ \\
F2V & 10.61 & $10.61 $ &  $9.95\pm 7.00$ & $10.47\pm 10.49$ & $ 10.36\pm $ 9.91\\
rG5IV & 16.67 & $16.67 $ & $16.83\pm3.46$ & $15.90\pm 6.47$ & $ 14.54\pm 9.93$ \\
rK3V & 15.15 & $15.15 $ &  $15.00\pm1.53$ & $14.66\pm 3.03$ & $ 14.85\pm 4.52$ \\
rM1V & 1.51 & $1.51 $ & $1.44\pm 0.35$ & $1.48\pm 0.73$ & $ 1.63\pm 1.49$ \\
G9III & 13.64 & $13.64 $ & $14.59\pm 6.28$ & $15.81\pm 11.31$ & $ 14.69\pm 11.93$ \\
rK3III & 12.12 & $12.12 $ & $11.91\pm 3.61$ & $11.84\pm 6.51$ & $ 12.82\pm 7.82$ \\
M5III & 3.03 & $3.03 $ & $2.69\pm 0.10$ & $2.72\pm 0.20$ & $ 3.06\pm 0.42$ \\
rG2I & 9.09 & $9.09 $ & $9.51\pm 3.11$ & $9.69\pm 5.86$ & $ 9.67\pm 9.22$ \\
M2I  & 4.54 & $4.55 $ & $4.45\pm 0.51$ & $4.37pm 1.04$ & $4.33 \pm 1.91$ \\
\hline
$D^2$ & -        & $4\, 10^{-16}$ &$0.086\pm 0.004$ & $0.344\pm 0.018$ & $1.387\pm 0.074$\\
\hline
E(B-V) & 0.2 & $0.2$ &$0.20\pm 0.01$ & $0.20\pm 0.02$ & $0.20\pm 0.03$ \\
\hline
$\sigma_{Dispersion}(km/s)$ &140  & $140$ & $140.8\pm 6.9$ & $142.0\pm 13.2$ & $144.0\pm 20.3$\\
\hline
\end{tabular}
\end{center}
\caption{Same as in Table~\ref{tab:sim1}, for a population dominated by intermediate-type stars.}
\label{tab:sim3}
\end{table*}

\begin{table*}[htbp]
\begin{center}
\begin{tabular}{|r|c|c|c|c|c|}
\hline
 & $\bm k_{Input} (\%)$ & $k_{Output} (\%)$ & $k_{Output} (\%)$ & $k_{Output} (\%)$ & $k_{Output} (\%)$ \\
     &                 & $\alpha=-1.5$   & $-0.5$    & $0.5$     & $1.5$\\
\hline
hot stars & $12$ & $19$ & $21$  & $28$ & $ 38$ \\
population &  inter &inter &inter  &inter & inter+young  \\
Continuum  & $10$ & - & - & - & - \\
\hline
$D^2$      & -    & $2\, 10^{-3}$ &$3\, 10^{-3}$ & $6\, 10^{-3}$ & $2\, 10^{-2}$\\
\hline
E(B-V)     & $0.0$ & $0.03$ &$0.07$ & $0.13$ & $0.23$ \\
\hline
$\sigma_{Dispersion}(km/s)$ & $0$ &$0$  & $0$ & $0 $ & $0 $\\
\hline
\end{tabular}
\end{center}
\caption{Solutions of a simulated galaxy dominated by an intermediate population with a contribution to luminosity of $12\%$ from hot stars and a $10\%$ contribution from a featureless continuum ($\bm k_{input}$). The different solutions correspond to the synthesis of the constructed galaxies where the additive continuum has a shape of a power law of different indices $\alpha$.}
\label{tab:simcont5}
\end{table*}

\begin{table*}[htbp]
\begin{center}
\begin{tabular}{|r|c|c|c|c|c|}
\hline
 & $\bm k_{Input} (\%)$ & $k_{Output} (\%)$ & $k_{Output} (\%)$ & $k_{Output} (\%)$ & $k_{Output} (\%)$ \\
     &                 & $\alpha=-1.5$   & $-0.5$    & $0.5$     & $1.5$\\
\hline
hot stars & $7$ & $46$ & $47$  & $36$ & $ 28$ \\
population & inter & young & young & inter+young & inter+young \\
Continuum  & $50$ & - & - & - & - \\
\hline
$D^2$      & -    & $3\, 10^{-2}$ &$4\, 10^{-2}$ & $0.22$ & $1.24$\\
\hline
E(B-V)     & $0.0$ & $0.17$ &$\geq 0.3$ & $\geq 0.3$ & $\geq 0.3$ \\
\hline
$\sigma_{Dispersion}(km/s)$ & $0$ &$\geq200 $  & $\geq 200$ & $\geq 200$ & $\geq 200$\\
\hline
\end{tabular}
\end{center}
\caption{Same as in Table~\ref{tab:simcont5}, but the contribution to luminosity of the additive continuum is of $50\%$.}
\label{tab:simcont6}
\end{table*}

\subsubsection{Results}\label{sec:resultssim}
The results of the Monte-Carlo simulations are shown in Tables \ref{tab:sim1} to \ref{tab:sim3} for three values of the Signal to Noise ratio (S/N $\sim$ 25, 50 and 100) and for three different populations : a population dominated by intermediate and late-type stars in Table \ref{tab:sim1}, by young and early-type stars in Table \ref{tab:sim2} and by intermediate stars in Table \ref{tab:sim3}.\\
These tables show clearly that for an infinite value of the S/N, the method provides the exact population. When the value of this ratio is larger than 50, the results are quite well-defined (with more than about $2\sigma$ error) in the example of the intermediate population (Table \ref{tab:sim3}). They are also well-determined for the contributions exceeding $7\%$ in the example of the intermediate and late-type population (Table \ref{tab:sim1}) and for contributions exceeding $9\%$ in the example of young early-type population (Table \ref{tab:sim2}).    

\subsection{Simulating an additive continuum}\label{sec:simcont}
In order to evaluate the influence of an additive continuum to the solutions obtained with this method, we constructed a set of galaxies without any reddening or velocity dispersion, in which $10\%$ of the V-band luminosity arises from a power law continuum with index $\alpha=-1.5,-0.5,0.5,1.5$. Another set of galaxies in which the contribution to luminosity of the continuum is of $50\%$ is also constructed. The synthesis procedure is then performed on the spectrum of the constructed galaxies. The test is made on three different populations as described in Sect.~\ref{sec:resultssim} (a population dominated by young and early-type stars, a second population dominated by intermediate-type stars and the last one dominated by late-type stars). We show the results for the intermediate population in Tables \ref{tab:simcont5} and \ref{tab:simcont6}.\\
The striking result is that in all cases the solutions show a large increase of the contributions to luminosity of hot (O- and A-type) stars. This result suggests that whatever the slope of the additive continuum (i.e. increasing or decreasing), the remedy of the absence of this continuum in the database is a high contribution of hot stars since these are the only spectral types showing a quasi-featureless continuum in the optical band. \\
The high contribution of hot O- and A-type stars results in higher values of the reddening, especially in the case of positive indices (increasing function) of the power law. This effect is expected, since the continuum shape of the hot stars is decreasing, therefore, to straighten up the spectra, a higher reddening is needed.\\
It is also noteworthy that for a contribution to luminosity of 50$\%$ of the additive continuum, high velocity dispersions are required as well to synthesise the spectrum (see Table~\ref{tab:simcont6}).\\
This test shows that the stellar population composition is highly affected by the presence of an additive continuum (e.g. the presence of an AGN or heated dust in the IR case) if this continuum is not modeled in the synthesis. In particular, a considerable artificial contribution to the luminosity of young hot population appears in such cases. \\
Moreover, a clear correlation appears between the reddening, the velocity dispersion and the presence of a featureless continuum. This correlation can be expected (but not easily proved) since the additive continuum changes the slope of the spectrum and partly affects the lines in a similar way to the velocity dispersion. Indeed, an additive continuum dilutes the absorption lines of a spectrum resulting in the reduction of the line strengths as is the case when the spectrum is smoothed.


\section{Application to the spectrum of the globular cluster G~170}\label{sec:application}
\begin{figure*}
\rotatebox{0}{
\resizebox{15cm}{!}{{\includegraphics{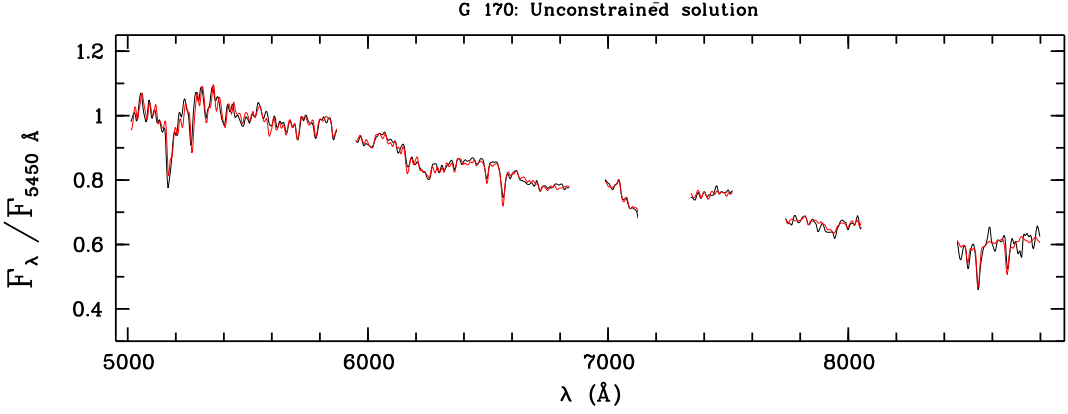}}}}  
\caption{Synthetic and observed spectra respectively in dark and light lines for the unconstrained solution of the globular cluster G~170.}
\label{G170Flux.ps}
\end{figure*}

\begin{figure}
\rotatebox{0}{
\resizebox{7cm}{!}{{\includegraphics{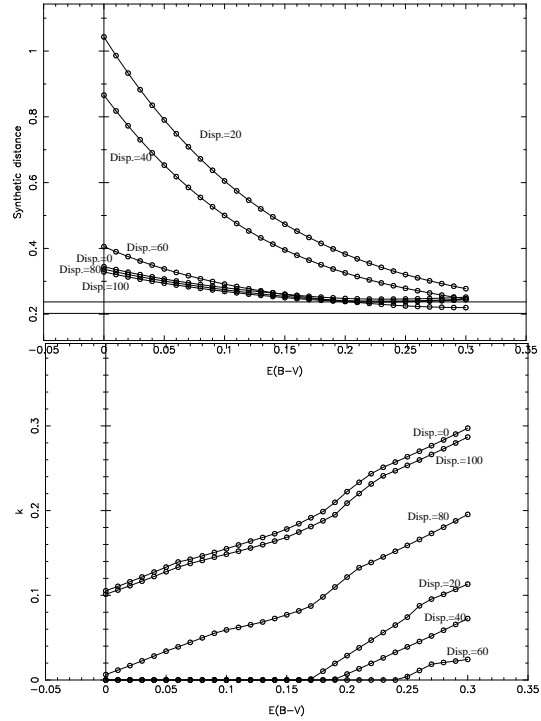}}}}  
\caption{Synthesis of the whole spectrum of the globular cluster G~170: Variation with reddening of the synthetic distance (top) and the contribution to luminosity of O- and A-type stars ($k$) (down) for different fixed dispersion values ($\sigma=0, 20, 40, 60, 80$ and $100 km/s$). The horizontal lines delimit the minimal synthetic distance with its $1 \sigma$ error.}
\label{red95.ps}
\end{figure}

\begin{figure*}
\rotatebox{0}{
\resizebox{15cm}{!}{{\includegraphics{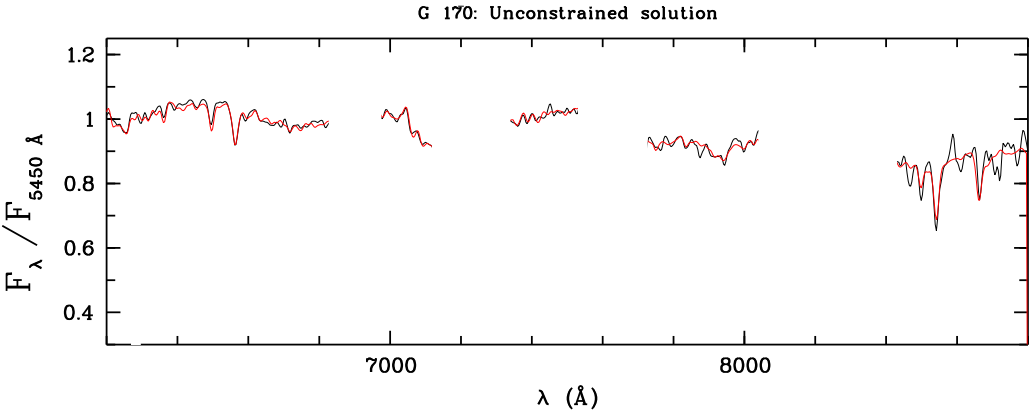}}}} 
\caption{Same plot as in Fig.~\ref{G170Flux.ps} but for the synthesis performed only with the red part of the spectrum (from 6200 to 9000$\AA$).} 
\label{G170Fluxredpart.ps}
\end{figure*}

\begin{figure}
\rotatebox{0}{
\resizebox{7cm}{!}{{\includegraphics{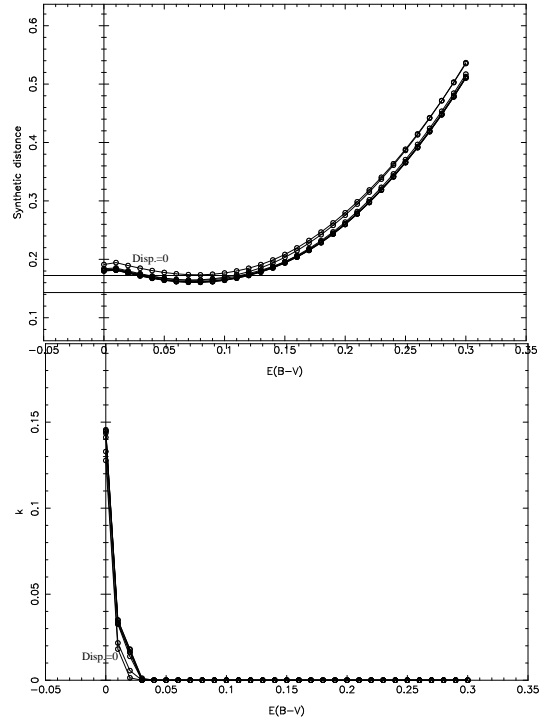}}}}  
\caption{Same plot as in Fig.~\ref{red95.ps} but corresponding to the synthesis of the red part of the spectrum of G~170 (from 6200 to 9000$\AA$).}
\label{red89.ps}
\end{figure}

The method is applied to the optical spectrum of the globular cluster G~170 of the galaxy M31; the spectral resolution is of $\Delta \lambda=17 \AA$. The database used for the synthesis is the one described in Boisson et al. (2000) with a spectral resolution of $11\AA$. The choice of this example is motivated by the fact that the stellar population of this cluster has been analysed in Moultaka et al. (2004) using the EW method and in Jablonka et al. (1992). It is thus an ideal example to study with the present method since we wish to compare the results with those of the EW method and of an independent analysis. For this reason, the choice of the stellar database is restricted to a low resolution database of 31 uncorrelated stars (at this spectral resolution and spectral range from $\sim 5000\AA$ to $\sim 9000\AA$, only this number of stars is necessary to represent the HR diagram, see Sect. \ref{sec:database}). Obviously, the present method can be applied to high resolution spectra of composite objects using stars from the new libraries such as the Indo-US library (Valdes et al. 2004), the UVES library (Bagnulo et al. 2003), the ELODIE library (Soubiran et al. 1998) or the ELODIE archive after a flux calibration of the spectra (Moultaka et al. 2004) and many others, or synthetic libraries as that of Bertone et al. (2004) etc... In that case, because of the higher spectral resolution, the computing time would increase since for the same wavelength range, a higher number of pixels would be present in the spectra and a larger number of stars would be used for the synthesis (because the number of uncorrelated stars would also increase).\\
The results of the synthesis obtained with the present method (the unconstrained and the constrained solutions with the ``Decreasing IMF'' and the ``Standard'' modes) are shown in Tables~\ref{G170synt} and \ref{G170syntredpart} and in Figs.~\ref{G170Flux.ps} and \ref{G170Fluxredpart.ps}. The results listed in Table~\ref{G170synt} are obtained using all the intensities of the spectrum except four regions which were not corrected for the atmospheric bands (see Boisson et al. 2000) and the region of the Na absorption line (5874 to 5914$\AA$) which is also affected by the ISM. The spectrum of the unconstrained solution is shown in Fig.~\ref{G170Flux.ps} superimposed on the dereddened observed spectrum. This figure shows that the spectrum is very well synthesised. However, the stellar population of this solution is not satisfactory because most of the contributions are not well determined suggesting that the stellar database is not adequate for the analysis of the globular cluster in that spectral range, spectral resolution and S/N ratio of the globular cluster spectrum. In addition, the contribution of hot A-type stars and the high reddening of $\sim 0.3$ mag are probably not real since this globular cluster is known to be $2\,10^9$ yrs old (Jablonka et al. 1992) and reddened by $\sim 0.05$ mag. For this reason, we searched for other solutions where different parts of the spectrum were not used in the synthesis procedure. We found that the contribution of hot stars disappears when using only the red part of the spectrum (for instance from $6200\AA$ to $9000\AA$). The results are shown in Table \ref{G170syntredpart}. The obtained unconstrained solution is well defined and provides a similar stellar population as the one obtained by Moultaka et al. (2004). It agrees with the known age of this cluster but the population here is metal rich. The resulting value of the velocity dispersion is due to the non comparable spectral resolution of the stellar database in comparison with the globular cluster spectral resolution. The synthetic and observed spectra are shown in Fig.~\ref{G170Fluxredpart.ps}. \\
 The need of hot quasi-featureless stars in the previous solution obtained using the whole spectrum is thus due to the blue part of the spectrum. Both unconstrained solutions have similar populations but the solution of the synthesis of the whole spectrum has in addition a contribution of A-type stars because a bluer continuum is needed in this case. This behaviour is well shown in Fig.~\ref{red95.ps} where the variations of the synthetic distance and the contribution from hot stars with reddening are plotted. For all dispersions, the contribution of hot stars increases with the reddenning while the synthetic distance decreases. The optimal solution is on a border of the domain of parameters (at E(B-V)=0.3). This suggests that the minimum is a local one but a global minimum is probably at a higher value of the reddening which is not acceptable in physical terms (for instance, the minimal solution is at E(B-V)=0.52 mag with a contribution of 22$\%$ of hot stars). On the other hand, a similar plot corresponding to the synthesis of the red part of the spectrum (see Fig.~\ref{red89.ps}) shows that in this case, the optimal solution lies inside the domain of parameters and is therefore more reliable. In Moultaka et al. (2004), the solution does not show a contribution from hot stars. This results from the use of an empirical continuum to deduce the values of the EWs of absorption lines. 
 The additional ``empirical'' information of the continuum used in the EWs method cannot be used in our case. A wider wavelength range is needed to provide the information about the blue part of the spectrum. This information is provided indirectly and empirically in the EWs method but is not justified as no information on the shape of the continuum in the blue part of the spectrum is given. The lack of information in the blue part of the spectrum is a weakness whatever the synthesis method is (direct or inverse). However, since in the present method the quality of the solution is also estimated thanks to the error analysis, it is possible to know the confidence level of the solution (in the present case for example, the solution was not well determined suggesting thus a lack of information).\\
This analysis shows once more that the choice of the database and the wavelength range are crucial in the synthesis procedure and that the wavelength range should harbour enough signatures of the different stellar classes. It should also allow one to distinguish between the stellar classes within the noise level of the composite object spectrum as explained in Sect. \ref{sec:database}.\\

For both syntheses, the constrained ``Decreasing'' mode solutions are equal or equivalent to the unconstrained solution as was the case for the method with EWs (Moultaka et al. 2004). The ``Decreasing IMF'' mode provides solutions which are acceptable in the sense that their synthetic distances lie inside the 1$\sigma$ error around the synthetic distance of the unconstrained solution. In Table \ref{G170syntredpart}, the solution provides an earlier turnoff of the stellar population (in G4V suggesting an age of $2\, 10^9$ yrs).\\
The results are very satisfactory since the unconstrained solutions are globally the same as the constrained ones. This shows that the method is very stable according to such constraints i.e. that the unconstrained solutions are physically acceptable.
 
\begin{table*}[htbp]
\small
\begin{center}
\begin{tabular}{rccc}
\hline
          & Unconstrained        & Dec. IMF    & Standard   \\
Star      & solution ($\%$)   & mode ($\%$)    & mode ($\%$)     \\
\hline
O7-BOV    & $0$          & $0$         & $0$ \\    
B3-4V     & $2\pm1.5$    & $2\pm1$     & $2\pm1.5 $ \\     
A1-3V     & $0$          & $0.06\pm0.3$& $0.2\pm0.002$  \\     
F2V       & $0$          & $0$         & $0$ \\
F8-9V     & $0$          & $0$         & $0$  \\
G4V       & $0$          & $4\pm5$     & $0$ \\   
G9-K0V    & $0$          & $0.6\pm8$   & $0$  \\  
K5V       & $5\pm19$     & $7.5\pm19$  & $3\pm19$ \\
M2V       & $0.3\pm1$    & $1\pm1$     & $0.1\pm1$  \\ 
rG0IV     & $0$          & $0$         & $0$ \\    
rG5IV     & $4\pm17$     & $8\pm13$    & $5\pm17$  \\   
rK0V      & $0$          & $0$         & $0$  \\ 
rK3V      & $10\pm8$     & $6\pm9$     & $9.5\pm8$   \\   
rM1V      & $0$          & $0$         & $0$  \\   
G0-4III   & $63\pm 3$    & $56\pm1$    & $64\pm3$  \\    
wG8III    & $0$          & $0$         & $0$  \\   
G9III     & $0$          & $0$         & $0$  \\   
K4III     & $0$          & $0$         & $0$  \\    
M0.5III   & $0$          & $0$         & $0$    \\  
M4III     & $5\pm1$      & $5\pm1$     & $5\pm1$ \\ 
M5III     & $0$          & $0$         & $0$  \\ 
rG9III    & $0$          & $0$         & $0$  \\   
rK3III    & $0$          & $0$         & $0$  \\   
rK3IIIbis & $7\pm12$     & $7\pm10$    & $7\pm12$   \\    
rK5III    & $3\pm13$     & $3\pm12$    & $3\pm13$    \\ 
G0Iab     & $0$          & $0$         & $0$    \\ 
K4Iab     & $0$          & $0$         & $0$   \\    
M2Ia      & $0$          & $0$         & $0$     \\  
rG2Iab    & $0$          & $0$         & $0$   \\   
rK0II     & $0$          & $0$         & $0$     \\   
rK3Iab    & $0$          & $0$         & $0$  \\   
\hline
$D^2$     & $0.22\pm0.017$  & $0.22\pm 0.017$ & $0.22\pm0.015$ \\
\hline
E(B-V)    & $0.30\pm0.10$       & $0.30\pm0.10$         & $0.31\pm0.1$   \\
Dispersion ($\sigma$)& $60\pm 340$ & $60\pm 340$  & $60\pm 340$\\
\hline
\end{tabular}
\end{center}
\caption{Results of the spectral synthesis of the globular cluster G~170. Each column displays the stellar contributions with their standard deviations for three solutions: the unconstrained solution and the solutions of the two modes of constraints described in the text. $D^2$ is the synthetic distance and E(B-V) the reddening. The dispersion is given in $km/s$.}
\label{G170synt}
\end{table*}

\begin{table*}[htbp]
\small
\begin{center}
\begin{tabular}{rccc}
\hline
          & Unconstrained        & Dec. IMF    & Standard   \\
Star      & solution ($\%$)   & mode ($\%$)    & mode ($\%$)     \\
\hline
O7-BOV    & $0$          & $0$         & $0$ \\    
B3-4V     & $0$          & $0$         & $0$ \\     
A1-3V     & $0$          & $0$         & $0$  \\     
F2V       & $0$          & $0$         & $0$ \\
F8-9V     & $0$          & $0$         & $0$  \\
G4V       & $0$          & $16\pm1.5$  & $0$ \\   
G9-K0V    & $0$          & $3\pm14$    & $0$  \\  
K5V       & $0$          & $4\pm11$    & $0$ \\
M2V       & $0$          & $1\pm0.4$   & $0$  \\ 
rG0IV     & $0$          & $0$         & $0$ \\    
rG5IV     & $0$          & $0$         & $0$  \\   
rK0V      & $29\pm2$     & $3\pm19$    & $29\pm2$  \\ 
rK3V      & $9\pm4$      & $0$         & $9\pm4$   \\   
rM1V      & $0$          & $0$         & $0$  \\   
G0-4III   & $12\pm12$    & $7\pm 1.5$  & $12\pm12$  \\    
wG8III    & $0$          & $0$         & $0$  \\   
G9III     & $0$          & $6\pm12$    & $0$  \\   
K4III     & $0$          & $0$         & $0$  \\    
M0.5III   & $0$          & $0$         & $0$    \\  
M4III     & $4\pm1$      & $4\pm0.5$   & $4\pm1$ \\ 
M5III     & $2\pm0.5$    & $2\pm0.5$   & $2\pm0.5$  \\ 
rG9III    & $45\pm6$     & $54\pm5$    & $45\pm6$  \\   
rK3III    & $0$          & $0$         & $0$  \\   
rK3IIIbis & $0$          & $0$         & $0$   \\    
rK5III    & $0$          & $0$         & $0$    \\ 
G0Iab     & $0$          & $0$         & $0$    \\ 
K4Iab     & $0$          & $0$         & $0$   \\    
M2Ia      & $0$          & $0$         & $0$     \\  
rG2Iab    & $0$          & $0$         & $0$   \\   
rK0II     & $0$          & $0$         & $0$     \\   
rK3Iab    & $0$          & $0$         & $0$  \\   
\hline
$D^2$     & $0.16\pm0.015$ & $0.17\pm 0.015$ & $0.16\pm0.015$ \\
\hline
E(B-V)    & $0.09\pm0.06$        & $0.08\pm0.03$         & $0.09\pm 0.06$   \\
Dispersion ($\sigma$)& $140\pm60 $ & $140\pm 60$  & $140\pm 60$\\
\hline
\end{tabular}
\end{center}
\caption{Synthesis solutions for the globular cluster G~170 obtained using only the red part of the spectrum (from $6200$ to $9000\AA$).}
\label{G170syntredpart}
\end{table*}

\section{Summary and conclusion}
In this paper, an inverse method of stellar population synthesis is described.
This method uses all intensities at each pixel of a composite object spectrum as observables to be fitted by a combination of spectra from a database.\\
The novelty of this work is that the reddening, velocity dispersion and dust emission in the IR part of the spectrum are also modeled and an analytical computation of the errors around a solution is provided. \\
Linear (astro)physical constraints can be included while searching for a solution. Constraints on the IMF have shown that the unconstrained solutions are very satisfactory as they satisfy such constraints.\\
Tests of the method have shown that for a S/N ratio larger than 50, the solution is well determined and when this ratio is infinite, the solution is the exact one.\\
Finally, in the present spectral optical band used in this paper, it has been shown that the synthesis problem is very sensitive in the blue part. The presence of hot stars in the solutions is often artifactual due to a lack of information in the blue part of the spectrum. The present spectral interval does not contain sufficient spectral signatures of hot stars and thus cannot constrain the presence of hot stars in the solution. For the same reason, the presence of a featureless continuum in a composite spectrum results in a high contribution of hot stars in the synthesis solution. 

\appendix 
\section{The synthetic surface and the gradients of functions $Q_r$:}
The ``synthetic surface'' is defined as the set of vectors of the normalised intensity vector space for which there exists a solution. In other words, the ``synthetic surface'' is the set of vectors $\bm I_{syn}$ satisfying the equation system:\\
\begin{equation}
\left\{
\begin{array}{ll}
\bm {II}\bm k-\bm I_{syn}= &\bm 0 \\
\bm b \bm k - 1= &0\\
\end{array}
\right.
\end{equation}

The previous system can be written as follows :
\begin{equation} 
\bm {II}^{++}\bm k^{+}=\bm 0 
\end{equation}
where $\bm {II}^{++}=\left(\begin{array}{c|c}\bm {II}^{+}&\bm I_{syn}\\
                                                         & 1\end{array}\right)$ and  
$\bm k^{+}=\left(\begin{array}{c}
\bm k\\
-1
\end{array}\right)$.\\
This equation is true only if for each combination of $n_\star+1$ lines of the matrix $\bm {II}^{++}$, the determinant of the resulting matrix equals zero (because vector $\bm k^{+}$ is never equal to zero). Thus, as the rank of matrix $\bm {II}^{++}$ is $n_\star$, one only needs to search for the $n_\star$ linearly independent lines and impose that the determinants of the matrices obtained using these lines and to which is added a remaining line $r$ ($\bm {II}^{++}_r$), are equal to zero. Let us call the obtained determinants (which are of a number of $n_l+1-n_\star$) $Q_r$ where $r$ is the index corresponding to the $n_l+1-n_\star$ remaining lines. It follows:\\
\begin{equation}
Q_r=det(\bm {II}^{++}_r)=\sum_{i=1}^{r}(-1)^{n_{\star}+1+i}I_{syn\,i} det(\bm {II}^{++\,i(n_{\star}+1)}_r)
\end{equation}
where $\bm {II}^{++\,i(n_{\star}+1)}_r$ is the extracted matrix from matrix $\bm {II}^{++}_r$ by eliminating the $i^{th}$ line and the last column (i.e. the $n_{\star}+1^{th}$ column, which is vector $\bm I_{syn}$).\\
In order to shift the origin of the vector space to $\bm I_{syn \,0}$, we perform a translation of vector $-\bm I_{syn \,0}$ to the matrices $\bm {II}^{++}_r$ and obtain:
\begin{equation}
\begin{array}{lll}
\frac{\partial Q_r}{\partial I_{syn\,j}}|_{I_{syn\,j0}} &=(-1)^{n_{\star} +1+j}det(\bm {II}^{++\,j(n_{\star}+1)}_r) &  \,\,\,\, if \,\,j=1,...,n_{\star},r\\
\frac{\partial Q_r}{\partial I_{syn\,j}}|_{I_{syn\,j0}} &= 0 \,\,\,\, otherwise.
\end{array}
\end{equation} 

\section{Facilities of the code:}
The Synthesis code is written in such a way that several entities can be comfortably changed or included. Here are listed some of them:
\begin{itemize}
\item All components which may compose the galactic spectrum consist of independent files containing the pixel values of the corresponding spectra. Thus, the choice of the presupposed composition of the studied spectrum is completely free. One may use all possible stellar, cluster and/or even galactic spectra. One may also use Blackbody spectra and/or any wavelength function.\\
This facility can be used to test for example additive continua in the case of AGNs.
\item The user of the code can choose wavelength intervals (from one pixel to the whole spectrum) to be excluded from the synthesis procedure. This is useful for example in the case of spectral regions showing emission lines or not well calibrated.
\item Different weights can be included easily in the input of the program for each of the used pixel intensities of the spectra. This may be used to give more or less importance to some spectral regions in the synthesis procedure. In this case, the synthetic distance has the following form:
\begin{equation}
\begin{array}{ll}
D^2   &=\sum_{j=1}^{n_l}(I'_{gal\,j}-I'_{syn\,j})^2
\end{array}
\end{equation}
where $I'_{gal\,j}$ and $I'_{syn\,j}$ are the weighed galactic and synthetic intensities ($I'_{gal\,j}=I_{gal\,j}\,\sqrt P_j$ and $I'_{syn\,j}=\,\sum_{i=1}^{n_{\star}} k_i I_{ji} \sqrt P_j$). 
\item Dust spectra can be included easily in the database like the stellar spectra. They are also constructed with the required starting wavelength value, pixel step and pixel number using a dedicated code. A subroutine of this code allows the user to decide the shape of the function describing the dust emission.
\item The reddening law may also be changed easily, it consists of an independent file listing the pixel values of the law in the wavelength domain.
\item The dispersion is tested by convolving the spectra of the database with a Gaussian of chosen parameters. Any other function as the model using the Gauss-Hermite series can be easily included as a subroutine in the code.
\item At present, the constraints have the form of large inequalities between the different numbers (or equivalently the different contributions to luminosity at the reference wavelength) of stars or objects of the database. Therefore, any linear constraints may be chosen in the code. 
\item The interval values for all parameters can be changed confortably. 
\end{itemize}
An independent code ``Proginterpol.f90'' allowing users to homogenise the different spectra (the galactic or stellar spectra or the spectra of dust, reddening or any wavelength function) is also available. This code rebins the spectra to a single step and makes them all begin (first pixel) at the same wavelength. It allows one as well to obtain spectra having the same number of~pixels.\\ 

\section{The output of the program:}
The code produces 4 files:
\begin{itemize} 
\item The ``solution'' which consists of a list of the contributions (and their standard deviations) of each object of the database, the reddening and the dispersion with their errors.  
In this file are also listed all the input parameters of the synthesis.
\item The observed spectrum dereddened with the best obtained value. The file consists on an ASCII file listing the pixel values of the spectrum.
\item The synthetic constructed spectrum resulting from the combination of the contributions of the objects in the database.
\item An information file listing all used parameters and for each iteration the synthetic distance, the reddening, the dispersion and the contributions of hot stars. This file is used to plot figures showing the behaviour of the synthetic distance and hot stars with the different parameters as Figs.~\ref{red95.ps} and \ref{red89.ps}. 
\end{itemize} 

\begin{acknowledgements}
I would like to thank A. Eckart for his support and C. Boisson, F. Boone, A. Eckart, M. Joly, D. Pelat and Ph. Prugniel for interesting discussions. I am grateful to A. Eckart and Ph. Prugniel for their comments.\\
This work was supported in part by the Deutsche Forschungsgemeinschaft (DFG) via grant SFB 494.
\end{acknowledgements}

\end{document}